\documentclass[10pt,a4paper]{article}
\usepackage[latin1]{inputenc}
\usepackage{appendix}
\usepackage[cmex10]{amsmath}
\usepackage{amssymb,amsthm,amsfonts}
\usepackage{amsmath}
\usepackage{amsfonts}
\usepackage{amssymb}
\usepackage{graphicx}
\usepackage{epstopdf}
\usepackage{url}
\theoremstyle{plain}
\newtheorem{theorem}{Theorem}
\newtheorem{lemma}{Lemma}

\newtheorem{definition}{Definition}
\theoremstyle{remark}
\newtheorem{rem}{Remark}
\newtheorem{example}{Example}
\newtheorem{cor}{Corollary}

\usepackage[left=3cm, right=2cm, top=3cm, bottom=3cm]{geometry}
\begin{document}

\title{Aliasing-truncation Errors in Sampling Approximations of Sub-Gaussian Signals}

\author{Yuriy Kozachenko and Andriy Olenko% <-this % stops a space
\thanks{Yu. Kozachenko is with the Department of Probability Theory, Statistics and Actuarial Mathematics, Kyiv University, Kyiv 01601, Ukraine (e-mail: ykoz@ukr.net)}% <-this % stops a space
\thanks{A. Olenko is with the School of Engineering and Mathematical Sciences, La Trobe University, Victoria 3086, Australia (e-mail: a.olenko@latrobe.edu.au)}
\thanks{This research was partially supported under Australian Research Council's Discovery Projects funding scheme (project number DP160101366) and  La Trobe University DRP Grant in Mathematical and Computing Sciences.  The authors are also grateful for the referees' careful reading of the paper and suggestions, which helped to improve the paper.}
\thanks{Manuscript received December 27, 2014; revised November 20, 2015 and June 16, 2016.}
\thanks{Copyright (c) 2014 IEEE. Personal use of this material is permitted.  However, permission to use this material for any other purposes must be obtained from the IEEE by sending a request to pubs-permissions@ieee.org}}

% The paper headers
\markboth{IEEE TRANSACTIONS ON INFORMATION THEORY,~Vol. , No. ,   2016}%
{Kozachenko and Olenko: Aliasing-truncation errors in approximation of
\mbox{$Sub_{\varphi}(\Omega)$} signals}

% make the title area
\maketitle

% As a general rule, do not put math, special symbols or citations
% in the abstract or keywords.
\begin{abstract}
The article starts with new aliasing-truncation error upper bounds in the sampling theorem for non-bandlimited stochastic signals.
Then, it investigates $L_p([0,T])$ approximations of sub-Gaussian random signals. Explicit truncation error upper bounds are established. The obtained rate of convergence provides a constructive algorithm for determining the sampling rate and the sample size in the truncated Whittaker-Kotel'nikov-Shannon expansions to ensure the approximation of sub-Gaussian signals with given accuracy  and reliability. Some numerical examples are presented.

\

\end{abstract}

% Note that keywords are not normally used for peerreview papers.

% For peer review papers, you can put extra information on the cover
% page as needed:
% \ifCLASSOPTIONpeerreview
% \begin{center} \bfseries EDICS Category: 3-BBND \end{center}
% \fi
%
% For peerreview papers, this IEEEtran command inserts a page break and
% creates the second title. It will be ignored for other modes.
\maketitle

\section{Introduction}
Recovering signals from discrete samples and estimating the information loss are the fundamental problems in sampling theory and signal analysis. Whittaker-Kotel'nikov-Shannon (WKS) theorem is a classical tool to recover a continuous band-limited signal from a sequence of its discrete samples. WKS theorems are extensively used in communications and information theory. Various new refine results are published regularly by engineering and mathematics communities, see, e.g., \cite{boch,but,fer,he,xue} and the recent volumes  \cite{hog,moe,zay}.

A growing body of work uses the sampling reconstruction of stochastic signals  to model various real physical processes. However, the sampling theory for the case of stochastic signals is much less developed comparing to its  deterministic counterpart. The publications  \cite{boch, fay, he, ole2, pog} and references therein  present an almost exhaustive survey of key approaches in stochastic sampling theory. 

The majority of known stochastic sampling results were obtained for harmonizable random processes.  Spectral representations of  these random processes were used to directly employ the known deterministic sampling results and error bounds for finding mean square approximation errors, see, e.g.,  \cite{he, ole1, ole2, pog, son} among other works. 
However, in applications,  various measures of the closeness of trajectories throughout  the  entire  signal  support are often more appropriate than mean-square errors in each time point. Controlling  signal distortions in the mean-square sense may result in situations where realizations of stochastic signals are substantially distorted. Instead of small mean-square errors in each location one may need to guarantee that the signal trajectories have not been changed more than a certain tolerance. For example,  near-lossless compression requires small user-defined tolerance levels, see [5,11]. It indicates the necessity of elaborating special stochastic techniques.

Another problem is that, in practice, a signal may not have a band-limited spectrum. Recently a considerable attention was given to this problem. One of possible approaches  is to use wavelet series representations of stochastic processes, see \cite{kozol1, kozol2}. Notice that the WKS sampling is an example of such general expansions but requires specific methods and techniques. Another approach to deal with non-bandlimited signals is to simultaneously increase the sampling rate and the sample size used in the truncated WKS formula.

The aliasing error appears due to the divergence between the actual band-region of the spectrum of a signal  $\mathbf X(t)$
and the assumed one.
The aliasing error is defined as the difference between the initial signal $\mathbf X(t)$
and its cardinal series expansion, reading as follows:
\begin{equation}\label{ali}
\mathbf X(t)-\sum_{k=-\infty}^\infty \frac{\sin\left(\omega\left(t-\frac{k\pi}{\omega}\right)\right)}{\omega\left(t-\frac{k\pi}{\omega}\right)}\, \mathbf X\left(\frac{k\pi}{\omega}\right).
\end{equation}
For stationary stochastic signals a sharp upper bound for the aliasing error was first obtained in \cite{bro} and further investigated in \cite{ole0,ole1}. 

However,
in contrast to this theoretical approach,  the infinite sum in (\ref{ali}) cannot be employed in numerical implementations as only finitely many samples
are available. Hence, one has to estimate the combined aliasing-truncation error
\[
\mathbf X(t)-\sum_{k=-n}^n \frac{\sin\left(\omega\left(t-\frac{k\pi}{\omega}\right)\right)}{\omega\left(t-\frac{k\pi}{\omega}\right)}\, \mathbf X\left(\frac{k\pi}{\omega}\right).
\]
\begin{figure}[!h]
\begin{center}
\includegraphics[width=8.5cm,height=6.9cm,trim=5mm 4mm 2mm 10mm,clip]{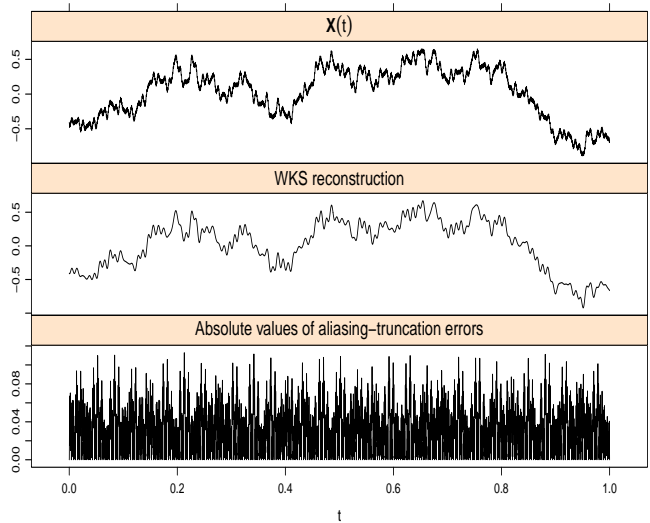}
\caption{Signal, its WKS reconstruction, and aliasing-truncation error.}
\label{fig:1}
\end{center}
\end{figure}

The analysis presented in the paper addresses the above problems and contributes to the former stochastic sampling literature.  New aliasing-truncation sampling results are derived for the class of sub-Gaussian random processes. This class  extends various properties of Gaussian processes to more general settings. The aliasing-truncation error for the WKS expansions has never been studied for sub-Gaussian random processes. Also, a thorough search of the relevant literature only yielded  sampling truncation and aliasing results for stochastic signals in the mean-square metric. There are no known truncation and aliasing results on   WKS approximations of trajectories of stochastic signals in $L_p([0,T])$ metrics in the literature. 

Note, that the first progress in studying sampling reconstructions of sub-Gaussian random signals was made in \cite{kozol0}. The authors investigated reconstruction errors of bandlimited signals in  $L_p([0,T])$ and uniform norms. This paper extends authors' methodology in \cite{kozol0} to non-bandlimited signals. For the non-bandlimited case it is not enough only to increase the sample size to decrease a reconstruction error. One has to increase both the sampling rate and the sample size. Also, in contrast to
the theoretical results in \cite{ole0,ole1}, the new approach makes it possible to obtain explicit upper bound estimates which can be used across a range of applications. 

The analysis also gives a constructive algorithm for determining sampling rates and  sample sizes to ensure the reconstruction of stochastic signals with a given accuracy.
The results can be useful in classical applications in  communications and information theory and new areas of compressed sensing, see, e.g., \cite{hog, xue}. 

The article proceeds as follows. The next section  introduces two classes of sub-Gaussian random processes. Section~\ref{sec3} derives new aliasing-truncation error upper bounds in the WKS approximation of non-bandlimited stochastic signals. Then,  Section~\ref{sec4} presents results on the approximation of $\varphi$-sub-Gaussian signals in  $L_p([0,T])$ with a given accuracy and reliability. Some applications of the obtained results are demonstrated in Section~\ref{sec5}. Finally, we conclude the paper with a short discussion and some problems for further investigation.

\section{Truncation error upper bounds for non-bandlimited processes}\label{sec3}
This section presents new truncation error upper bounds in the WKS approximation of non-bandlimited stochastic processes.  To the best of the authors' knowledge the combined aliasing-truncation error has not been studied for stochastic processes, except the weak Cram\'{e}r class, see \cite{ole1}. However,  the upper bounds in \cite{ole1} were investigated under additional decay conditions  on the family of functions determining stochastic processes. Also these results were not in the explicit form ready for numerical implementations. 

 Let $\mathbf X(t),$ $t\in \mathbf{R},$ be a stationary real-valued mean square continuous 
random process with $ \mathbf E\mathbf \mathbf X(t)=0.$ The process $\mathbf X(t)$ yields the spectral representation 
\[\mathbf X(t)= \int_{-\infty}^\infty e^{it \lambda} {\rm d}\eta(\lambda),\]
where $\eta(\cdot)$ is a random process with uncorrelated increments.
Then, its covariance function is given by
\[\mathbf B(\tau):=\mathbf E\mathbf X(t+\tau) \mathbf X(t)= \int_{-\infty}^\infty e^{i\tau \lambda} {\rm d}F(\lambda),\] 
where $F(\cdot)$ is the spectral function of $\mathbf X(t)$ such that $\mathbf E\left[ \eta(b)-\eta(a)\right]^2=F(b)-F(a),$ $b\ge a.$

Let us define the corresponding process $ \mathbf X_\Lambda(t)$ 
whose spectrum is bandlimited to $[-\Lambda,\Lambda]$ as follows
\[\mathbf X_\Lambda(t)= \int_{-\Lambda}^\Lambda e^{it \lambda} {\rm d}\eta(\lambda).\]

Then, for all $\omega>\Lambda$ there holds
\begin{equation}\label{shan1}
\mathbf X_\Lambda(t)=\sum_{k=-\infty}^\infty \frac{\sin\left(\omega\left(t-\frac{k\pi}{\omega}\right)\right)}{\omega\left(t-\frac{k\pi}{\omega}\right)}\, \mathbf X_\Lambda\left(\frac{k\pi}{\omega}\right),
\end{equation}
and the series (\ref{shan1}) converges uniformly in mean square.

Let us consider the truncation versions given by the formulae
\begin{equation}\label{shan20}
\mathbf X_n(t):=\sum_{k=-n}^n \frac{\sin\left(\omega\left(t-\frac{k\pi}{\omega}\right)\right)}{\omega\left(t-\frac{k\pi}{\omega}\right)}\, \mathbf X\left(\frac{k\pi}{\omega}\right),
\end{equation}
\begin{equation}\label{shan2}
\mathbf X_{\Lambda,n}(t):=\sum_{k=-n}^n \frac{\sin\left(\omega\left(t-\frac{k\pi}{\omega}\right)\right)}{\omega\left(t-\frac{k\pi}{\omega}\right)}\, \mathbf X_\Lambda\left(\frac{k\pi}{\omega}\right).
\end{equation}

The following result gives a new upper bound for aliasing-truncation errors in the mean square norm.
 \begin{theorem}\label{the0}  Let $z\in (0,1),$ $t >0,$ and  $n \ge \frac{\omega t}{\pi \sqrt{z}}.$ Then  
\[\left(\mathbf E \left|\mathbf X(t)-\mathbf X_n(t)\right|^2\right)^{1/2}\le A_n(t,\Lambda) ,\]
where 
\[A_n(t,\Lambda):= \frac{\sqrt{C_n(t,\Lambda)}}{n}+D_n(t)\cdot\left(\int_{|\lambda|>\Lambda}{\rm d}F(\lambda)\right)^{1/2},\]
\[C_n(t,\Lambda):=\mathbf B(0)\cdot\left(\frac{4\omega t}{\pi^2 (1-z)}+\frac{4\left(z+1+\frac{1}{n}\right)}{\pi (1-z)^2\left(1-\frac{\Lambda}{\omega}\right)} \right)^2,
\]
\[D_n(t):=2+ 
\frac{2\left|\sin(\omega t)\right|}{\pi}\left(1+\mbox{\rm Si}(\pi)+\frac{\frac{\omega t}{\pi}+0.5}{n-\frac{\omega t}{\pi}}\right).\]
 \end{theorem} 
 For fixed $\omega,$ $t,$ and $z$ the inequality $n \ge \frac{\omega t}{\pi \sqrt{z}}$ provides the sufficient sample size $n$ to guarantee that the mean square reconstruction errors do not exceed the specified level $A_n(t,\Lambda).$ 
 
\begin{rem}\label{rem1} Note that $C_n(t,\Lambda)$ is bounded by $C_1(t,\Lambda)$ and  $D_n(t)\le 2+ 
\frac{2\left|\sin(\omega t)\right|}{\pi}\left(1+\mbox{\rm Si}(\pi)+\frac{\frac{\omega t}{\pi}+0.5}{1-\sqrt{z}}\right).$  
Therefore, for fixed $t$, $\omega,$ and $z,$  there is some constant $C,$ for example,

 $C=\max\left(C_1(t,\Lambda),2+ 
 \frac{2\left|\sin(\omega t)\right|}{\pi}\left(1+\mbox{\rm Si}(\pi)+\frac{\frac{\omega t}{\pi}+0.5}{1-\sqrt{z}}\right)\right),
 $ such that
 \[A_n(t,\Lambda)\le C\cdot\left(n^{-1}+\left(\int_{|\lambda|>\Lambda}{\rm d}F(\lambda)\right)^{1/2}\right).\]
  
Hence,  $A_n(t,\Lambda)\to 0, $ when both $n$ and $\Lambda$ go to $+\infty.$
\end{rem}
  
\begin{cor}\label{cor0}  Let $z^*=\frac{\omega^2 T^2}{\pi^2 n^2}\in (0,1),$ $T >0.$ Then  
\[\sup_{t\in [0,T]}\left(\mathbf E \left|\mathbf X(t)-\mathbf X_n(t)\right|^2\right)^{1/2}\le \tilde{A}_n(T,\Lambda),\]
where 
\[\tilde{A}_n (T,\Lambda):= \frac{\sqrt{\tilde{C}_n(T,\Lambda)}}{n}+\tilde{D}_n(T)\cdot\left(\int_{|\lambda|>\Lambda}{\rm d}F(\lambda)\right)^{1/2},\] 
\[\tilde{C}_n(T,\Lambda):=\mathbf B(0)\cdot\left(\frac{4\omega T}{\pi^2 (1-z^*)}+\frac{4\left(z^*+1+\frac{1}{n}\right)}{\pi (1-z^*)^2\left(1-\frac{\Lambda}{\omega}\right)} \right)^2,
\]
\[\tilde{D}_n(T):=2+ 
\frac{2}{\pi}\left(1+\mbox{\rm Si}(\pi)+\frac{\frac{\omega T}{\pi}+0.5}{n-\frac{\omega T}{\pi}}\right).\]
 \end{cor}

 \section{$\varphi$-Sub-Gaussian random processes}\label{sec2}
 This section reviews the definition of $\varphi$-sub-Gaussian random processes and  their relevant properties. 
 
Numerous applications use stationary Gaussian processes. This is justified by the central limit theorem where a resulting process is obtained as a sum of a large number of processes with small variances. However, if summands have relatively large variances the central limit theorem is not applicable and a resulting process may not be Gaussian. At the same time,  $\varphi$-sub-Gaussianity often is still a plausible assumption. For example, all centered bounded processes are $\varphi$-sub-Gaussian. Also, sums of independent Gaussian and centered bounded processes are  $\varphi$-sub-Gaussian.  Moreover,  under very mild conditions, the processes admitting  the Karhunen-Lo\'{e}ve type expansion are strictly $\varphi$-sub-Gaussian.  
 
 The space of $\varphi$-sub-Gaussian random variables  was introduced in \cite{kozost} to generalize sub-Gaussian results from  \cite{kah} to more general settings. Tail distributions of sub-Gaussian random variables behave similar to the Gaussian ones so that sample path properties of sub-Gaussian processes rely on their mean square regularity. Various properties of $\varphi$-sub-Gaussian random variables were studied in the book  \cite{bul} and the articles  \cite{kozost, ant1}. More information on $\varphi$-sub-Gaussian random processes and their applications can be found in the publications  \cite{bul, ant1,  fer} and references therein.
 
  \begin{definition}  A continuous even convex function $\varphi(x),$ $x\in {\mathbb R},$ is called
 an Orlicz N-function, if it is monotonically increasing for $x>0$,
 $\varphi(0)=0,$ $ {{\varphi(x)}/{x}}\to 0,$ when $x\to 0,$ and $ {{\varphi(x)}/{x}}\to\infty,$  when $x\to\infty. $
  \end{definition}
 
  \begin{definition} Let $\varphi(x),x\in {\mathbb R},$ be an Orlicz N-function. The
 function $\varphi^{*}(x):=\sup_{y\in {\mathbb R}}(xy-\varphi(y)),$ $x\in {\mathbb R},$ is called the
 Young-Fenchel transform of $\varphi(\cdot).$  \end{definition}

  \begin{definition} An Orlicz N-function  $\varphi(\cdot)$ satisfies {\bf Condition~Q}~if
 \[\lim\limits_{x\rightarrow 0}{\varphi(x)}/{x^2}=C>0,\]
 where the constant $C$ can be equal to $+\infty.$ 
  \end{definition}
 
  \begin{lemma}\label{lem1}  Let  $\varphi(\cdot)$  be an Orlicz N-function. Then it can be represented as
 $\varphi(u)=\int_{0}^{|u|}f(v)\,dv,$ where $f(\cdot)$ is
 a monotonically nondecreasing, right-continuous function, such that
 $f(0)=0$ and $f(x)\to +\infty,$ when  $x\to +\infty.$   \end{lemma}
 
 Let $\{\Omega, \cal{B}, P\}$ be a standard probability space and $L_p(\Omega)$ denote a space of random variables having finite $p$-th absolute moments.
 
  \begin{definition} Let $\varphi(\cdot)$ be an Orlicz
 N-function satisfying the {Condition Q.} A zero mean random
 variable $\xi$ belongs to the space $Sub_\varphi(\Omega)$ (the space
 of $\varphi$-sub-Gaussian random variables), if there exists a
 constant $a_{\xi}\geq 0$ such that the inequality $\mathbf
 E\exp\left(\lambda\xi\right)\le
 \exp\left(\varphi(a_{\xi}\lambda)\right)$ holds for all $\lambda\in
 {\mathbb R}.$  \end{definition}
 The space $Sub_\varphi(\Omega)$ is equipped
 the norm (see  \cite{bul}) 
 $$
 \tau_\varphi \left( \xi \right): = \mathop {\sup }\limits_{\lambda
 \ne 0} \frac{\varphi ^{\left( { - 1} \right)}\left( {\ln \mathbf
 E\exp
 \left\{ {\lambda \xi } \right\}} \right)}{\left| \lambda
 \right|},
 $$
 where $\varphi ^{\left( { - 1} \right)}(\cdot)$ denotes the inverse function of $\varphi (\cdot).$
 
 If  $\varphi(x)={x^2}/2$ then the space $Sub_\varphi(\Omega)$ is called a space of sub-Gaussian variables. It was introduced in the article~\cite{kah}.

  \begin{definition} Let $\mathbf{T}$ be a parametric space. A random process $\mathbf
 X(t),$ $t \in \mathbf{T},$ belongs to the space $Sub_{\varphi}(\Omega)$ if $\mathbf X(t)\in Sub_{\varphi}(\Omega)$ for
 all $t\in \mathbf{T}.$ 
  \end{definition}
 Gaussian centered random process $\mathbf X(t),$ $t\in \mathbf{T},$ belongs to the space $Sub_{\varphi}(\Omega),$  where $\varphi(x)={x^2}/2$ and  $\tau_{\varphi}(\mathbf X(t)) =\left(\mathbf
 E\left|\mathbf X(t)\right|^2\right)^{1/2}.$  If $\mathbf X(t)$ is a centered bounded random variable for all $t\in \mathbf{T}$ then the process  $\mathbf X(t),$ $t\in \mathbf{T},$ belongs to all spaces $Sub_{\varphi}(\Omega).$      
Another example is the case when $\mathbf X(t)$ is a two-sided Weibull random variable,  i.e. 
 \[\mathbf P\left\{{ \mathbf X(t)}\ge x \right\}=\mathbf P\left\{ {\mathbf X}(t) \le -x \right\}=\frac{1}{2}\exp\left\{-\frac{x^\alpha}{\alpha}\right\}, \quad x>0.\]  
  Then  $\mathbf X(t),$ $t\in \mathbf{T},$ is a random process from the space $Sub_{\varphi}(\Omega).$ 
 
  \begin{definition}\label{SSub}  A family $\Xi$ of random
 variables $\xi\in Sub_{\varphi}(\Omega)$ is called strictly
 $Sub_{\varphi}(\Omega)$ if there exists a constant $C_{\Xi}>0$
 such that for any finite set $I,$ $\xi_{i}\in \Xi,$ $i\in I,$
 and for arbitrary $\lambda_{i}\in {\mathbb R},$ $i\in I:$ 
 \[\tau_{\varphi}\left(\sum\limits_{i \in
 I}\lambda_{i}\xi_{i}\right)\leq
  C_{\Xi}\left(\mathbf E\left(\sum\limits_{i \in I}\lambda_{i}\xi_{i}\right)^2\right)^{1/2}.\]
   \end{definition}
 $C_{\Xi}$ is called a determinative constant.
 The strictly $Sub_{\varphi}(\Omega)$ family will be denoted by $SSub_{\varphi}(\Omega).$ 
 
  \begin{definition} $\varphi$-sub-Gaussian random
 process $\mathbf X(t),$ $t\in \mathbf{T},$ is called strictly
 $Sub_{\varphi}(\Omega)$ if the family of random variables $\{\mathbf
 X(t),t\in \mathbf{T}\}$ is strictly $Sub_{\varphi}(\Omega).$ The determinative
 constant of this family is called a determinative constant of the
 process and denoted by $C_{\mathbf X}$.
  \end{definition}
 
 A Gaussian centered random process $\mathbf X(t),$ $t\in \mathbf{T},$ is a 
 $SSub_{\varphi}(\Omega)$ process, where $\varphi(x)={x^2}/2$ and the
 determinative constant $C_{\mathbf X}=1.$ 
 
\section{Approximation in $L_p([0,T])$}\label{sec4}
This section presents novel results on the WKS approximation of non-bandlimited sub-Gaussian signals in  $L_p([0,T]),$  $p \ge 1,$ with a given accuracy and reliability.  Notice, that the approximation in $L_p([0,T])$ provides the closeness of trajectories of the signal $\mathbf X(t)$ and its approximant $\mathbf X_n(t),$ see, e.g.,  \cite{ kozol1, kozol2}. It is different from the known $L_p$-norm results (in particular the mean square results in Section~\ref{sec3}) which give only the closeness of distributions for each $t,$ see, e.g.,  \cite{he, ole1, ole0}.

 \begin{theorem}\label{the3}  Let $\mathbf X(t),$ $t\in \mathbf{R},$ be a stationary
$SSub_{\varphi}(\Omega)$ process. 
Let  $\mathbf X_n(t)$ be defined by {\rm(\ref{shan20})}, $\Lambda\in(0,\omega).$ 

Then,  $\int_0^T
\left|\mathbf X(t)-\mathbf X_n(t)\right|^p\,dt$ exists with probability~{\rm 1} and  for any $\varepsilon>S_{n,p}\cdot\left(f\left(p\,(S_{n,p}/\varepsilon)^{1/p}\right)\right)^p$ the following inequality holds true
\[\mathbf P\left\{\int_0^T
\left|\mathbf
X(t)-\mathbf
X_n(t)\right|^p\,dt>\varepsilon\right\}\le 2\exp\left\{-\varphi^*\left(\left({\varepsilon}/S_{n,p}\right)^{1/p}\right)\right\},
\]
where  
\[S_{n,p}:=C_{\mathbf X}^p\int_0^T A_n^{p}(t,\Lambda)\,dt\]
 $C_{\mathbf X}$ is a determinative constant of the process  $\mathbf X(t),$  $f$ is from Lemma~{\rm\ref{lem1}}, and $A_n(t,\Lambda)$ was defined in Theorem~{\rm\ref{the0}. }

 \end{theorem}
\begin{rem}\label{rem2} It can be seen from Remark~\ref{rem1} that, for fixed $T,$ $\omega,$ and $z,$  we obtain
 \[S_{n,p}\le \tilde{C}\cdot\left(n^{-1}+\left(\int_{|\lambda|>\Lambda}{\rm d}F(\lambda)\right)^{1/2}\right)^p,\]
where, for example, $\tilde{C} =C_{\mathbf X}^p\cdot C^p\cdot T$ and the constant $C$ is defined in Remark~\ref{rem1}.

Hence,  $S_{n,p}\to 0, $ when both $n$ and $\Lambda$ go to $+\infty.$
  \end{rem}
  
By Remark~\ref{rem2} and Lemma~\ref{lem1} the right hand side of the inequality $\varepsilon>S_{n,p}\cdot\left(f\left(p\,(S_{n,p}/\varepsilon)^{1/p}\right)\right)^p$ in Theorem~\ref{the3} is a decreasing function  of $n$ and $\Lambda.$ Hence, for each specific $\varepsilon,$ it gives the sufficient sample size $n$ to guarantee that Theorem~\ref{the3} holds true.

 \begin{example} Let $\varphi(x)={|x|^{\alpha}}/\alpha,$ $
1<\alpha\le2.$ Then $f(x)=x^{\alpha-1}$ and
$\varphi^*(x)={|x|^{\gamma}}/\gamma,$ where $\gamma \ge 2$ and $1/{\alpha}+1/{\gamma}=1.$ Hence,
 the condition of the theorem can be written as
\[\varepsilon>S_{n,p}\cdot\left(f\left(p(S_{n,p}/\varepsilon)^{1/p}\right)\right)^p=S_{n,p}^{\alpha}\,p^{(\alpha-1)p}\varepsilon^{1-\alpha}.\]
Therefore, it holds 
\[\mathbf P\left\{\int_\mathbf{T}
\left|\mathbf
X(t)-\mathbf
X_n(t)\right|^p\,dt>\varepsilon\right\}\le2\exp\left\{-\frac1\gamma\left(\frac{\varepsilon}{
S_{n,p}}\right)^{\gamma/p}\right\},\]
when
$\varepsilon>S_{n,p}\cdot p^{\frac{\alpha-1}{\alpha}p}.$

Remark~\ref{rem2} gives a constructive way to select $n$ and $\Lambda$ for a specific tolerance $\varepsilon.$
 \end{example}
Recalling that in the Gaussian case $\varphi^*(x)=|x|^2/2$ we obtain the following specification of Theorem~\ref{the3}.

\begin{example}  If $\mathbf X(t),$ $t\in \mathbf{R},$ is a Gaussian process, then for $\varepsilon>\hat{S}_{n,p}\cdot p^{p/2}$ it holds
\[\mathbf P\left\{\int_0^T
\left|\mathbf
X(t)-\mathbf
X_n(t)\right|^p\,dt>\varepsilon\right\}\le 2\exp\left\{-\frac{1}{2}\left(\frac{\varepsilon}{\hat{S}_{n,p}}\right)^{2/p}\right\},
\]
where \[\hat{S}_{n,p}:=\int_0^T A_n^{p}(t,\Lambda)\,dt.\]
\end{example}

 \begin{definition}\label{def8}  $\mathbf X_n(t)$ approximates $\mathbf X(t)$  in $L_p([0,T])$ with accuracy $\varepsilon>0$ and reliability $1-\delta,$ $0<\delta < 1,$ if 
\[\mathbf P\left\{\int_0^T \left|\mathbf X(t)-\mathbf X_n(t)\right|^p\,dt >\varepsilon\right\}\le \delta.\]
 \end{definition}

Using Definition~\ref{def8}  and Theorem~\ref{the3} we get the following result.
\begin{cor} Let $\mathbf X(t),$ $t\in \mathbf{R},$ be a stationary
$SSub_{\varphi}(\Omega)$ process. Then $\mathbf X_n(t)$  approximates $\mathbf X(t)$  in $L_p([0,T])$ with accuracy $\varepsilon$ and reliability $1-\delta$ if the following inequalities hold true
\[\varepsilon>S_{n,p}\cdot\left(f\left(p\,(S_{n,p}/\varepsilon)^{1/p}\right)\right)^p, 
\]
\[
 \exp\left\{-\varphi^*\left(\left({\varepsilon}/S_{n,p}\right)^{1/p}\right)\right\}\le \delta/2.
\]
 \end{cor}

\begin{cor}\label{cor2}  If $\mathbf X(t),$ $t\in \mathbf{R},$ is a Gaussian process, $\mathbf X_n(t)$  approximates $\mathbf X(t)$  in $L_p([0,T])$ with accuracy $\varepsilon$ and reliability $1-\delta$ if
\begin{equation}\label{Snp}\hat{S}_{n,p}<\frac{\varepsilon}{\max\left(p^{p/2},\left(2\ln(2/\delta)\right)^{p/2}\right)}.\end{equation}
\end{cor}

\section{Applications}\label{sec5}
This section discusses some practical steps  towards applications of the above theoretical results and  determining the number of terms in the WKS expansion for specified accuracy and reliability.

The straightforward analysis of the expressions  $\tilde{A}_n(T,\Lambda),$ $\tilde{C}_n(T,\Lambda),$ and $\tilde{D}_n(T)$ shows that  $\tilde{A}_n(T,\Lambda)\to 0$ when   $\Lambda$ and $n$ increase to $+\infty$ in such a way that $\Lambda/n \to 0$ and $\Lambda/\omega<1-\rho$ for some constant $\rho \in (0,1).$

The next example demonstrates that the upper bounds in Theorem~\ref{the0} and Corollary~\ref{cor0}  provide a reasonable approximation when the sampling rate and  sample size are sufficiently large. 

\begin{example}\label{ex3} For simulation studies in this example we use signals from the Whittle-Mat\'{e}rn family, see \cite[Section 5.2]{gel}. Namely, $\mathbf{X}(t)$ has the Whittle-Mat\'{e}rn covariance function  with the parameters $\alpha>0,$ $\phi =1,$ and $\nu=1/2$ if 
\[\mathbf{B}(\tau)=\sqrt{\frac{{2\pi\tau}}{{\alpha}}}\, K_{\frac{1}{2}}(\alpha\tau),\]
where $K_{\nu}(\cdot)$ is a modified Bessel function of the second kind.

Figure~\ref{fig:1} illustrates the WKS reconstruction of this stochastic signal. A simulated realization of the signal $\mathbf{X}(t),$ $t\in [0,1],$ is shown in the upper subplot.  The WKS reconstruction $\mathbf{X}_n(t)$ and the absolute value of aliasing-truncation errors $|\mathbf{X}(t)-\mathbf{X}_n(t)|$ are presented in the lower subplots.

Notice that the spectral density of the signal $\mathbf{X}(t)$ is
\begin{equation}\label{spec}\frac{{\rm d}F(\lambda)}{{\rm d}\lambda} = \frac{1}{\alpha^2+\lambda^2},\ \lambda \in \mathbf{R}.\end{equation}
Therefore, the spectrum of the signal $\mathbf{X}(t)$ is non-bandlimited. 

The first subplot of Figure~\ref{fig:2} demonstrates that for a fixed  $\Lambda$ the aliasing-truncation errors are substantial even for large values of $n.$ However, if we increase both $\Lambda$ and $n$ (as it is suggested below) the errors magnitude quickly decreases, see the second subplot in Figure~\ref{fig:2}. In the both subplots the same  $n$ was used.
\begin{figure}[!h]
\begin{center}
\includegraphics[width=9cm,height=7cm,trim=6mm 4mm 2mm 10mm,clip]{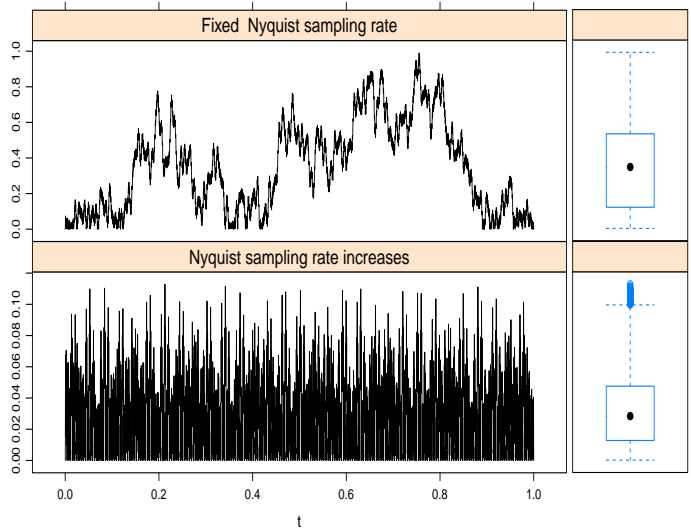}
\caption{Absolute values of aliasing-truncation errors}
\label{fig:2}
\end{center}
\end{figure}
 
Thus, contrary to the case of bandlimited signals, it is not enough to choose a sufficiently large $n$ to get an approximation with given accuracy and reliability. One needs to increase both the number of terms  $n$ and the sampling rate $2\Lambda$ simultaneously.

For simplicity, in this numerical example we consider $\omega=2^N$ and $\Lambda=3\omega/4,$ i.e. we double the sampling rate by increasing $N$ by 1. For  step $N+1$ this choice of sampling rate results in a refined grid  formed by dividing the interval between sample values used on step $N.$ Thus, on  step $N+1$ one can also use the sampled values from  step $N.$  We also select $n=N^2\cdot 2^N.$ Hence, $n,\Lambda \to +\infty$ and  $\Lambda/n \to 0$ when $N\to +\infty.$

Figure~\ref{fig:5} demonstrates the difference between the upper bound  $\tilde{A}_n(T,\Lambda)$ and Monte Carlo estimates of\linebreak $\sup_{t\in [0,T]}\left(\mathbf E \left|\mathbf X(t)-\mathbf X_n(t)\right|^2\right)^{1/2}$ obtained by simulating 50 realizations of $\mathbf{X}(t)$ for each $N$.  It is clearly seen that the upper bound  $\tilde{A}_n(T,\Lambda)$ approaches the supremum of mean square aliasing-truncation errors when $N$ increases.
\begin{figure}[!h]
\begin{center}
\includegraphics[width=9cm,height=5cm,trim=0mm 5mm 2mm 1.5cm,clip]{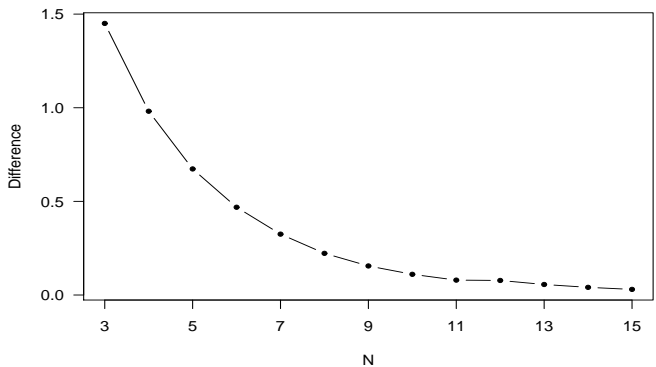}
\caption{Difference between $\tilde{A}_n(T,\Lambda)$ and the true value.}
\label{fig:5}
\end{center}
\end{figure}
\end{example}

Now we illustrate an application of the results in Section~\ref{sec4}  and determine the number of terms in the WKS expansion to ensure the approximation of $\varphi$-sub-Gaussian processes with given accuracy and reliability.

Let $p \ge 1$ in Corollary~{\rm\ref{cor2}}. Then, by Corollary~{\rm\ref{cor0}}, for arbitrary $z\in (0,1)$ and $n \ge \frac{\omega T}{\pi \sqrt{z}},$ we get the following estimate
\[\hat{S}_{n,p}\le \int_0^T A_n^p(t,\Lambda) dt\le  T\tilde{A}_n^p (T,\Lambda).\]

Hence, to guarantee {\rm(\ref{Snp})} for given $p,$ $\varepsilon$ and $\delta$ it is enough to choose such $n$ and $\Lambda$ that the following inequality holds true
\begin{equation}\label{N} \tilde{A}_n (T,\Lambda)\le \frac{(\varepsilon/T)^{1/p}}{\sqrt{\max\left(p,2\ln(2/\delta)\right)}}.
\end{equation}

To apply the obtained results for a specific class of stochastic signals one also needs to estimate the tail behaviour of the spectral function $F(\cdot).$  A widely used statistical approach is based on Abelian and Tauberian theorems, see \cite{leo} and references therein. It relies on the fact that the tail behaviour of the spectral function can be found from a local specification of  the covariance function in a neighbourhood of zero.

\begin{example}
Let $T=B(0)=1,$ and $p=2.$  Similar to Example~\ref{ex3} we choose $\omega=2^N,$  $\Lambda=3\omega/4,$ and $n=N^2\cdot 2^N.$ 
In this example we assume that $\int_{|\lambda|>\Lambda}{\rm d}F(\lambda)\le \Lambda^{-1}.$ For instance, the signals with spectral densities given by (\ref{spec}) satisfy this condition.  

The value of $N$ computed by (\ref{N}) as a function of  $\varepsilon$ and $\delta$ is shown in Figure~{\rm\ref{fig:3}}. It is clear that $n$ increases when $\varepsilon$ and $\delta$ approach~0, but for reasonably small $\varepsilon$ and $\delta$ we do not need too many sampled values.
\begin{figure}[!htb]
\begin{center}
\includegraphics[width=8.5cm,height=7cm,trim=1cmm 0.5cm 2cm 1.5cm,clip]{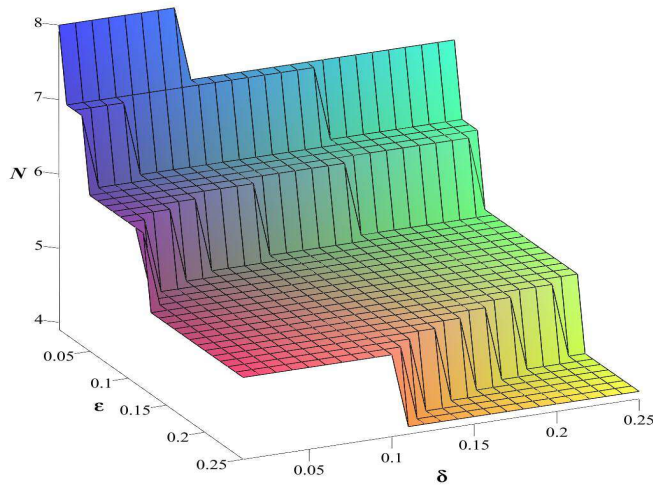}
\caption{The value of $N$ to ensure specified accuracy and reliability.}
\label{fig:3}
\end{center}
\end{figure}

For specific values of $\varepsilon,$ $\delta,$ and $p$ the obtained formulae give the sufficient sample size $N$ to guarantee that Theorems~\ref{the0} and \ref{the3} hold true. For example, for fixed $\varepsilon=\delta=0.1,$ Figure~{\rm\ref{fig:4}} illustrates the behaviour of the sufficient sample size as a function of the parameter~$p\in(1,2].$  
\begin{figure}[!htb]
\begin{center}
\includegraphics[width=7.5cm,height=6cm,trim=1mm 1mm 1mm 1mm,clip]{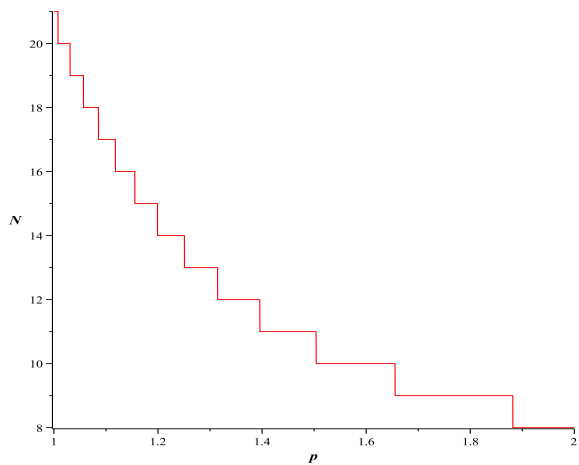}
\caption{The value of $N$ as a function of $p.$}
\label{fig:4}
\end{center}
\end{figure}
\end{example}

\section{Conclusions}
These results may have various applications to approximation problems  in signal processing and information theory. The obtained rate of convergence provides a constructive algorithm for determining the sample size and the sampling rate in the WKS expansions to ensure the approximation of $\varphi$-sub-Gaussian signals with given accuracy  and reliability. 
The developed methodology and new estimates are important extensions of the known results in the  stochastic sampling  theory to the space $L_p([0,T])$ and the class of $\varphi$-sub-Gaussian random processes. 

It would be of interest 
\begin{itemize}
\item to apply this methodology to other WKS sampling problems, for example, shifted sampling, irregular  sampling, see \cite{ole1,ole0,ole2} and references therein; 
\item to derive analogous results for the multidimensional case and spatial random processes;
\item to use approaches in \cite{kozol0} to obtain analogous results in the uniform metric. 
\end{itemize}

\newpage

\appendix
\appendixpage
\section{Proof of Theorem~\ref{the0}}

It follows from (\ref{shan1}), (\ref{shan20}), and (\ref{shan2}) that $\mathbf X(t)-\mathbf X_n(t)$ admits the following representation
\begin{eqnarray}
\mathbf X(t)&-&\mathbf X_n(t)=(\mathbf X(t)-\mathbf X_\Lambda(t))+(\mathbf X_\Lambda(t)-\mathbf X_{_\Lambda,n}(t))\nonumber\\
&+&(\mathbf X_{\Lambda,n}(t)-\mathbf X_n(t))=\int_{|\lambda|>\Lambda} e^{it \lambda} {\rm d}\eta(\lambda)\nonumber\\ 
&+&\sum_{|k|\ge n+1} \frac{\sin\left(\omega\left(t-\frac{k\pi}{\omega}\right)\right)}{\omega\left(t-\frac{k\pi}{\omega}\right)}\, \mathbf X_\Lambda\left(\frac{k\pi}{\omega}\right)\nonumber\\
&+&\sum_{k=-n}^n \frac{\sin\left(\omega\left(t-\frac{k\pi}{\omega}\right)\right)}{\omega\left(t-\frac{k\pi}{\omega}\right)}\left( \mathbf X_\Lambda\left(\frac{k\pi}{\omega}\right)- \mathbf X\left(\frac{k\pi}{\omega}\right)\right)\nonumber\\
&=&\int_{-\Lambda}^{\Lambda}\sum_{|k|\ge n+1} e^{i\lambda\frac{k\pi}{\omega}}\,\frac{\sin\left(\omega\left(t-\frac{k\pi}{\omega}\right)\right)}{\omega\left(t-\frac{k\pi}{\omega}\right)}\, {\rm d}\eta(\lambda)\nonumber\\
&+&\int_{|\lambda|>\Lambda} \left(e^{it \lambda} -\sum_{k=-n}^n e^{i\lambda\frac{k\pi}{\omega}}\,\frac{\sin\left(\omega\left(t-\frac{k\pi}{\omega}\right)\right)}{\omega\left(t-\frac{k\pi}{\omega}\right)}\right) {\rm d}\eta(\lambda).\nonumber
\end{eqnarray}

Therefore,
\[\left(\mathbf E \left|\mathbf X(t)-\mathbf X_n(t)\right|^2\right)^{1/2}\le  I_1+I_2\]
\[=
\left(\int_{-\Lambda}^{\Lambda}\left|\sum_{|k|\ge n+1} e^{i\lambda\frac{k\pi}{\omega}}\,\frac{\sin\left(\omega\left(t-\frac{k\pi}{\omega}\right)\right)}{\omega\left(t-\frac{k\pi}{\omega}\right)}\right|^2\,{\rm d}F(\lambda)\right)^{1/2}\]
\[ +
 \left(\int_{|\lambda|>\Lambda}\left|e^{i\lambda t}-\sum_{k=-n}^n e^{i\lambda\frac{k\pi}{\omega}}\,\frac{\sin\left(\omega\left(t-\frac{k\pi}{\omega}\right)\right)}{\omega\left(t-\frac{k\pi}{\omega}\right)}\right|^2 {\rm d}F(\lambda)\right)^{1/2}.\]

By Theorem 1 in \cite{kozol0} the first term $I_1$ can be bounded as
\begin{equation}\label{A1}
I_1^2\le \frac{C_n(t,\Lambda)}{n^2}.
\end{equation}
To estimate the second term $I_2$ we use the following bound
\[\left|e^{i\lambda t}-\sum_{k=-n}^n e^{i\lambda\frac{k\pi}{\omega}}\,\frac{\sin\left(\omega\left(t-\frac{k\pi}{\omega}\right)\right)}{\omega\left(t-\frac{k\pi}{\omega}\right)}\right|\le 1+|\sin\left(\omega t\right)|\]
\[\times
 \left|\sum_{k=-n}^n \frac{e^{ik\pi\left(\frac{\lambda}{\omega}+1\right)}}{\omega\left(t-\frac{k\pi}{\omega}\right)}\right|=:1+Z_n.\]
Let $k_t:=\mbox{nint}\left(\frac{tw}{\pi}\right),$ where $\mbox{nint}(x)$ denotes the integer closest to $x$ (a half integer rounds up), $a:=k_t-\frac{\omega t}{\pi}.$ Note that $k_t\ge 0$ and $|a|\le 0.5.$ Then,
\begin{eqnarray}
Z_n &\le& \left|\frac{\sin\left(k_t\pi-\omega t\right)}{k_t\pi-\omega t}\right|+
|\sin\left(\omega t\right)| \cdot\left|\sum_{\substack{k=-n \\k\not= k_t}}^n \frac{e^{ik\pi\left(\frac{\lambda}{\omega}+1\right)}}{k\pi-\omega t}\right|\nonumber\\
&\le & 1+ 
\frac{\left|\sin(\omega t)\right|}{\pi}\cdot\left|\sum_{\substack{k=-n-k_t \\k\not= 0}}^{n-k_t} \frac{e^{ik\pi\left(\frac{\lambda}{\omega}+1\right)}}{k+a}\right|\nonumber\\
&\le &  1 +
\frac{\left|\sin(\omega t)\right|}{\pi}\left(\left|\sum_{\substack{k=-n+k_t \\k\not= 0}}^{n-k_t} \frac{e^{ik\pi\left(\frac{\lambda}{\omega}+1\right)}}{k+a}\right|\right.\nonumber\\
& &{} +\left.\left|\sum_{k=n-k_t+1}^{n+k_t} \frac{1}{k-0.5}\right|\right)\nonumber\\
\end{eqnarray}
By $\frac{1}{k+a}=\frac{1}{k}\left(1-\frac{a}{k+a}\right)$ it follows that
\begin{eqnarray}
Z_n &\le & 1+ 
\frac{\left|\sin(\omega t)\right|}{\pi}\cdot\left(\frac{2|k_t|}{n-|k_t|+0.5}\right.\nonumber\\
& &{}\left.+\left|\sum_{\substack{k=-n+k_t \\k\not= 0}}^{n-k_t} \frac{e^{ik\pi\left(\frac{\lambda}{\omega}+1\right)}}{k}\left(1-\frac{a}{k+a}\right) \right| \right)\nonumber\\
&\le & 1+ 
\frac{2\left|\sin(\omega t)\right|}{\pi}\left(\left|\sum_{k=1}^{n-k_t} \frac{\sin\left(k\pi\left(\frac{\lambda}{\omega}+1\right)\right)}{k}\right|\right.\nonumber\\
&  &{}+\left.\frac{|k_t|}{n-|k_t|+0.5}+\left|
\sum_{\substack{k=-n+k_t \\k\not= 0}}^{n-k_t} \frac{a}{2k\left(k+a\right)}\right|\right).
\nonumber
\end{eqnarray}
The last sum above can  be estimated as follows
\begin{eqnarray}
& &\left|\sum_{\substack{k=-n+k_t \\k\not= 0}}^{n-k_t} \frac{a}{2k\left(k+a\right)}\right|= \frac{|a|}{2}\left| \sum_{k=1}^{n-k_t} \left(\frac{1}{k\left(k+a\right)}\right.\right.\nonumber\\
& &\left.\left.+\frac{1}{k\left(k-a\right)}\right)\right|
= \frac{1}{2}\left| \sum_{k=1}^{n-k_t} \left(\frac{1}{k} -\frac{1}{k+a}+\frac{1}{k-a}-\frac{1}{k}\right)\right|
\nonumber\\
& &\le \frac{1}{2}\left| \sum_{k=1}^{n-k_t} \left(\frac{1}{k-0.5} -\frac{1}{k+0.5}\right)\right|\le 1.
\nonumber
\end{eqnarray}
Note that 
\[\frac{|k_t|}{n-|k_t|+0.5}\le  \frac{\frac{\omega t}{\pi}+0.5}{n-\frac{\omega t}{\pi}}.\]
By  \cite[(3.5.5)]{mit} for all $n\in \mathbb N$ and $C\in \mathbb R$ \[\left|\sum_{k=1}^n\frac{\sin(Ck)}{k}\right|\le \int_0^\pi\frac{\sin(x)}{x}{\rm d}x=\mbox{Si}(\pi)\approx 1.8519.\]
Therefore we obtain
\[Z_n\le 1+
\frac{2\left|\sin(\omega t)\right|}{\pi}\left(1+\mbox{Si}(\pi)+\frac{\frac{\omega t}{\pi}+0.5}{n-\frac{\omega t}{\pi}}\right).
\]
Hence,
\begin{equation}\label{A2}
I_2^2\le (D_n(t))^2\cdot\int_{|\lambda|>\Lambda}{\rm d}F(\lambda).
\end{equation}
Now, the statement of the theorem follows from (\ref{A1}) and (\ref{A2}). 

% you can choose not to have a title for an appendix
% if you want by leaving the argument blank
\section{Proof of Theorem~\ref{the3}}
In this theorem we investigate the case when  $\mathbf T=[0,T]$ and $\mu$ is the Lebesgue measure  on $[0,T].$ Notice, that it follows from (\ref{shan2}) and Definition~\ref{SSub} that $\mathbf X(t)-\mathbf X_n(t)$ is a $SSub_{\varphi}(\Omega)$ random process with  the determinative constant~$C_{\mathbf X}.$ 

Now, we use the following result from {\rm \cite{koz0}}.

 \begin{lemma}\label{the1}  Let $p\ge 1$ and  
\[c:=\int_\mathbf{T}
\left(\tau_{\varphi}(t)\right)^p\,d\mu(t)<\infty.\]  Then the integral $\int_\mathbf{T}
\left|\mathbf X(t)\right|^p\,d\mu(t)$ exists with probability {\rm 1} and the following
inequality holds

\begin{equation}\label{est}\mathbf P\left\{\int_\mathbf{T}
\left|\mathbf
X(t)\right|^p\,d\mu(t)>\varepsilon\right\}\le 2\exp\left\{-\varphi^*\left(\left({\varepsilon}/c\right)^{1/p}\right)\right\}
\end{equation}
for each non-negative
\begin{equation}\label{epsilon_ner}\varepsilon>c\cdot\left(f\left(p(c/\varepsilon)^{1/p}\right)\right)^p,\end{equation}
where $f(\cdot)$ is a density of $\varphi(\cdot)$ defined in  Lemma~{\rm\ref{lem1}.}
 \end{lemma}

By the application of Lemma~\ref{the1} to $\mathbf X(t)-\mathbf X_n(t)$  we obtain that the integral $\int_0^T
\left|\mathbf X(t)-\mathbf X_n(t)\right|^p\,dt$ exists with probability~1.  In addition,
\[\mathbf P\left\{\int_0^T
\left|\mathbf
X(t)-\mathbf
X_n(t)\right|^p\,dt>\varepsilon\right\}\le 2\exp\left\{-\varphi^*\left(\left({\varepsilon}/c\right)^{1/p}\right)\right\},
\]
where $c:=\int_0^T \left(\tau_{\varphi}(\mathbf
X(t)-\mathbf
X_n(t))\right)^p\,dt.$ 

The functions $\varphi^*\left(\cdot\right)$ and $f(\cdot)$ are  monotonically  non-decreasing. Therefore, due to  $S_{n,p}\ge c,$ we obtain
\[S_{n,p}\cdot\left(f\left(p(S_{n,p}/\varepsilon)^{1/p}\right)\right)^p\ge c\cdot\left(f\left(p(c/\varepsilon)^{1/p}\right)\right)^p,\]
\[ \exp\left\{-\varphi^*\left(\left({\varepsilon}/c\right)^{1/p}\right)\right\}\le \exp\left\{-\varphi^*\left(\left({\varepsilon}/S_{n,p}\right)^{1/p}\right)\right\}.
\]

Hence, the statement of Lemma~\ref{the1} holds true if the constant $c$   is replaced by  $S_{n,p},$  which finishes the proof.

\section*{Supplementary Materials}
The codes used for simulations and examples in this article are available in the folder "Research materials" from \url{https://sites.google.com/site/olenkoandriy/}.
% use section* for acknowledgement

% Can use something like this to put references on a page
% by themselves when using endfloat and the captionsoff option.

\end{document}